# Self-emulsification in chemical and pharmaceutical technologies

D. Cholakova, Z. Vinarov, S. Tcholakova, N. Denkov*

*Department of Chemical and Pharmaceutical Engineering*
*Faculty of Chemistry and Pharmacy, Sofia University,*
*1 James Bourchier Avenue, 1164 Sofia, Bulgaria*

\*Corresponding author:
Prof. Nikolai Denkov
Department of Chemical and Pharmaceutical Engineering
Sofia University
1 James Bourchier Ave.,
Sofia 1164
Bulgaria
E-mail: nd@lcpe.uni-sofia.bg
Tel: +359 2 8161639
Fax: +359 2 9625643

# Abstract

The interest in the low energy self-emulsification techniques has exploded in the recent years, driven by three main trends: by the transition to "greener" technologies in both its aspects – less energy consumption and replacement of the petrochemicals by natural ingredients; by the costly and maintenance demanding equipment for nanoemulsification; and by the quest for efficient and robust self-emulsifying formulations for oral drug delivery. Here we first present a brief overview of the main known low-energy methods for nanoemulsion formation, focusing on their mechanistic understanding and discussing some recent advances in their development and applications. Next, we review three conceptually new approaches for self-emulsification in chemical technologies, discovered in the last several years. The colloidal features and the specific requirements of the self-emulsifying drug-delivery systems (SEDDS) are also discussed briefly. Finally, we summarize the current trends and the main challenges in this vivid research area.

## 1. Macroemulsions, nanoemulsions, microemulsions and emulsification approaches.

Emulsions are disperse systems of two immiscible liquids with different polarity. They are used in many consumer products, including cosmetics, pharmaceuticals, foods, inks, paints, agrochemicals, household and personal care products, and find important applications in various technologies, including chemical technologies, enhanced oil recovery, metalworking, firefighting, and others [1-5]. Due to the unfavorable contact between such phases with different polarity, called for brevity "water" and "oil", amphiphilic surfactants are added to decrease the interfacial energy and to stabilize the emulsion drops.

Emulsions, in which the drops are larger than 1 μm, are denoted in the literature as macroemulsions or just emulsions [1-6]. They are thermodynamically unstable because the Gibbs energy of the disperse system is higher than the Gibbs energy of the separated bulk phases:

$$\Delta G = \sigma \Delta A - T \Delta S \qquad \text{(eq. 1)}$$

Here $\Delta$ denotes the difference in the Gibbs energy, $G$, interfacial area, $A$, and entropy, $S$, of the emulsified and non-emulsified systems, respectively, at given temperature, $T$, and interfacial tension, $\sigma$. In macroemulsions, the increase of interfacial energy is larger in magnitude than the gain in entropy, $\Delta G > 0$, and mechanical energy is needed for emulsion formation [3,4].

Emulsions with drops in the nanometer scale have several benefits which make them preferred for practical applications. The gravitational separation (creaming/sedimentation), which is difficult to be suppressed in macroemulsions is no longer an issue for the emulsions containing nanodroplets, because the Brownian force and the natural convection (driven by small temperature gradients) become comparable in magnitude to the gravitational force. In addition, when the droplets become considerably smaller than the wavelength of the visible light, *ca.* < 50 nm, the emulsions become optically transparent which is preferred for application e.g. in refreshing beverages and sometimes in cosmetics [5,6]. In the realm of drug delivery, the oral bioavailability of lipophilic drugs increases significantly upon their administration as self-emulsifying (*viz.* creating nanodroplets) lipid-based formulations, compared to the lipid systems containing micrometer drops [7,8]. However, these advantages are at the expense of significantly larger surface area of the nanosized systems, which requires higher energy input to create the lipid drops and higher surfactant concentrations to stabilize them.

Two qualitatively different types of emulsion contain nanometer drops – microemulsions and nanoemulsions [6]. Microemulsions are thermodynamically stable, $\Delta G < 0$, whereas the nanoemulsions are thermodynamically unstable, **Figure 1a,b**. The thermodynamic stability of the microemulsions is related to their very low interfacial tension, $\sigma < 0.1$ mN/m, and the optimum curvature of the surfactant adsorption layers covering the microemulsion drops [6,9]. In contrast, the interfacial tension in the nanoemulsions is higher and $\Delta G > 0$.

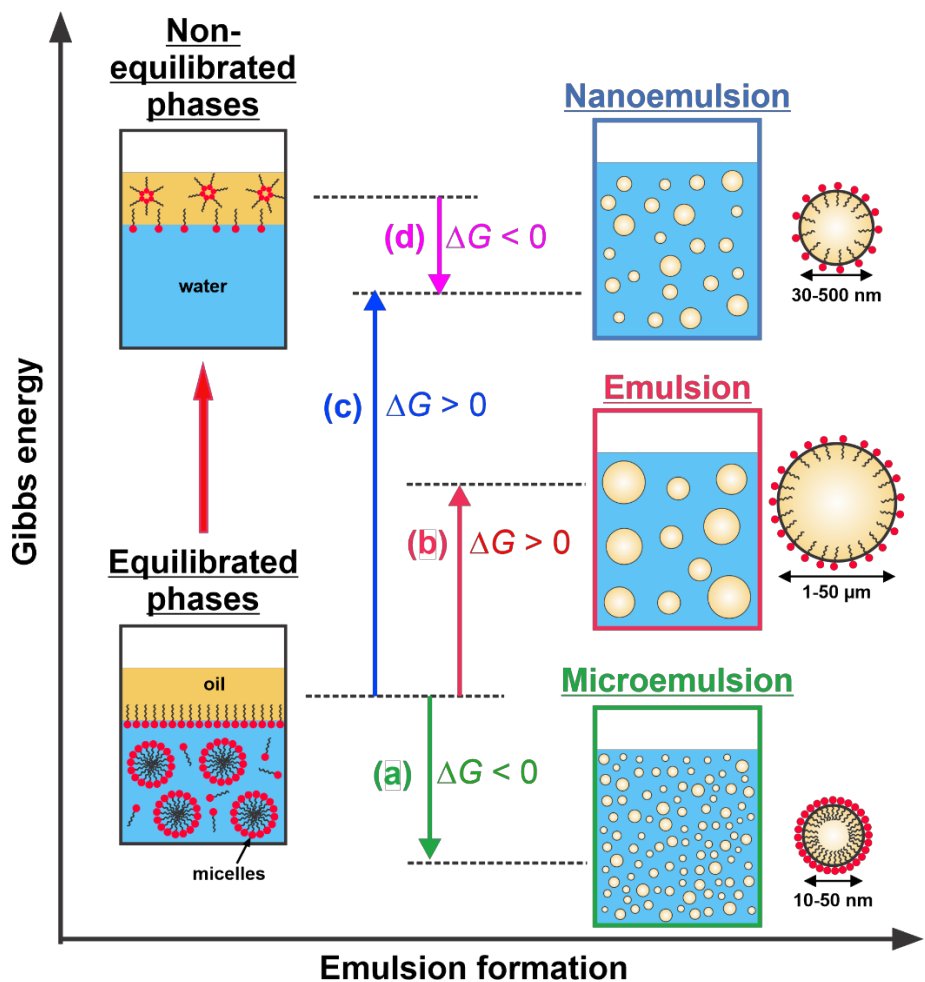

**Figure 1.** Schematic presentation of the change in the Gibbs energy upon formation of: (a) microemulsion; (b) macroemulsion; (c) nanoemulsion from equilibrated bulk phases and (d) nanoemulsion from non-equilibrated separated phases via self-emulsification. The surfactant micelles in the bulk phases are not shown in scale – they are much smaller than the droplets in the macro- and nanoemulsions and comparable in size to the microemulsion droplets.

Microemulsions can be considered as a kind of swollen micelles. The surfactant-to-oil ratio (SOR) in these systems is relatively high, because a comparable number of surfactant molecules are needed to solubilize the oil molecules in the swollen micelles. The attempt to increase the oil volume fraction in microemulsions leads to separation of a bulk oily phase at a certain critical value of the SOR. The microemulsions are indefinitely stable under given conditions, whereas different destabilization processes, such as Ostwald ripening, coalescence, flocculation and creaming occur upon storage of nanoemulsions and macroemulsions.

The microemulsion formation is a spontaneous process, although some energy input is often used to accelerate the incorporation of the oily molecules into the surfactant micelles [6]. In contrast, the process of formation of macroemulsions or nanoemulsions is not spontaneous, requires extra energy and is denoted in the literature as emulsification (oil-water mixing). The various emulsification methods are usually classified into two main groups, depending on the energy required for emulsion formation: high-energy methods and low-energy methods.

The high-energy emulsification, also called "top-down fabrication", is the most common approach used to induce drop breakage in the processes of macroemulsion and nanoemulsion formation. The equipment used in the high-energy methods includes high-pressure homogenizers, ultrasonicators, microfluidizers, rotor-stator homogenizers, such as colloidal mills, high shear mixers, toothed disk dispersing machines and others. All these methods do not require very specific selection of the materials, but they involve specialized, relatively expensive equipment with demanding maintenance. Another drawback of this approach is that usually > 99.9% of the energy is lost as heat and sound, due to the high friction in the respective devices, especially when nanoemulsions are formed – i.e. a very tiny fraction of the introduced energy is used for the actual drop breakage [3,10]. Recent reviews on these methods can be found in Refs. [5,11,12].

The low-energy methods are usually denoted as "self-emulsification" in the literature. The formation of nanoemulsions and macroemulsions is inherently an energy demanding process if we start with pre-equilibrated bulk oil and water phases, **Figure 1b,c**. However, when non-equilibrated bulk phases are used as starting materials, the process of emulsification can occur spontaneously and appears as self-emulsification, see **Figure 1d**.

In the following section 2, we present a brief overview of the main low-energy methods for nanoemulsion formation, known for years, with focus on their mechanistic understanding and some recent advances. Several novel approaches for self-emulsification, discovered in the last several years and with potential applications in the chemical and pharmaceutical technologies, are reviewed in Section 3. The specific requirements of the pharmaceutical drug-delivery systems are discussed briefly in Section 4. In the final Section 5, we summarize our outlook for the future development and the main challenges in this vivid research area.

## 2. Low-energy methods for (nano) emulsification

The methods for low-energy emulsification utilize the energy released upon phase transitions, surfactant transfer between different phases or other non-equilibrium processes which lead to the formation of nano-sized droplets. The most widely used methods are summarized in **Figure 2** and are described briefly in the current section.

### *Self-emulsification without surfactants*

The terms "Ouzo" or "Pastis" method, "emulsification by solvent diffusion (ESD)" and "solvent displacement method (SDM)" are used in the literature to describe processes of spontaneous emulsification which are caused by solubility changes of a solute upon addition of a second phase (in which the solute is less soluble) in the absence of surfactant [2,13-16]. This process was observed initially with the Greek alcoholic beverage Ouzo (called "Pastis" in France) [14]. Upon water addition, spontaneous phase separation occurs in the form of nano- or macroemulsion of the hydrophobic aromatic substance trans-anethole ($\approx$ 0.1%), which has been pre-dissolved in the original alcoholic beverage (with $\approx$ 45 % ethanol). Anethole is soluble in ethanol-water mixtures with high ethanol content, while it is poorly soluble in water. When the molecular anethole solution in 45 % ethanol (the Ouzo) is diluted with water, the solubility of the hydrophobic trans-anethole decreases and the anethole molecules nucleate spontaneously, forming small oily droplets inside the ethanol-water mixture.

This spontaneous emulsification process is diffusion driven and has been described by the diffusion path theory [13,17]. Particularly useful for identification of the phase behavior of such systems are the equilibrium ternary phase diagrams in which the binodal and spinodal curves enclose the Ouzo and Reverse Ouzo regions, **Figure 2a** [14]. Two main stages have been identified in the surfactant free emulsification by the Ouzo effect: initial extremely fast and extensive drop nucleation followed by a second stage of slow molecular transfer and possible drop coalescence [14,17]. Miniemulsion droplets as small as 10 nm have been produced via the Ouzo process under well controlled conditions [1,18].

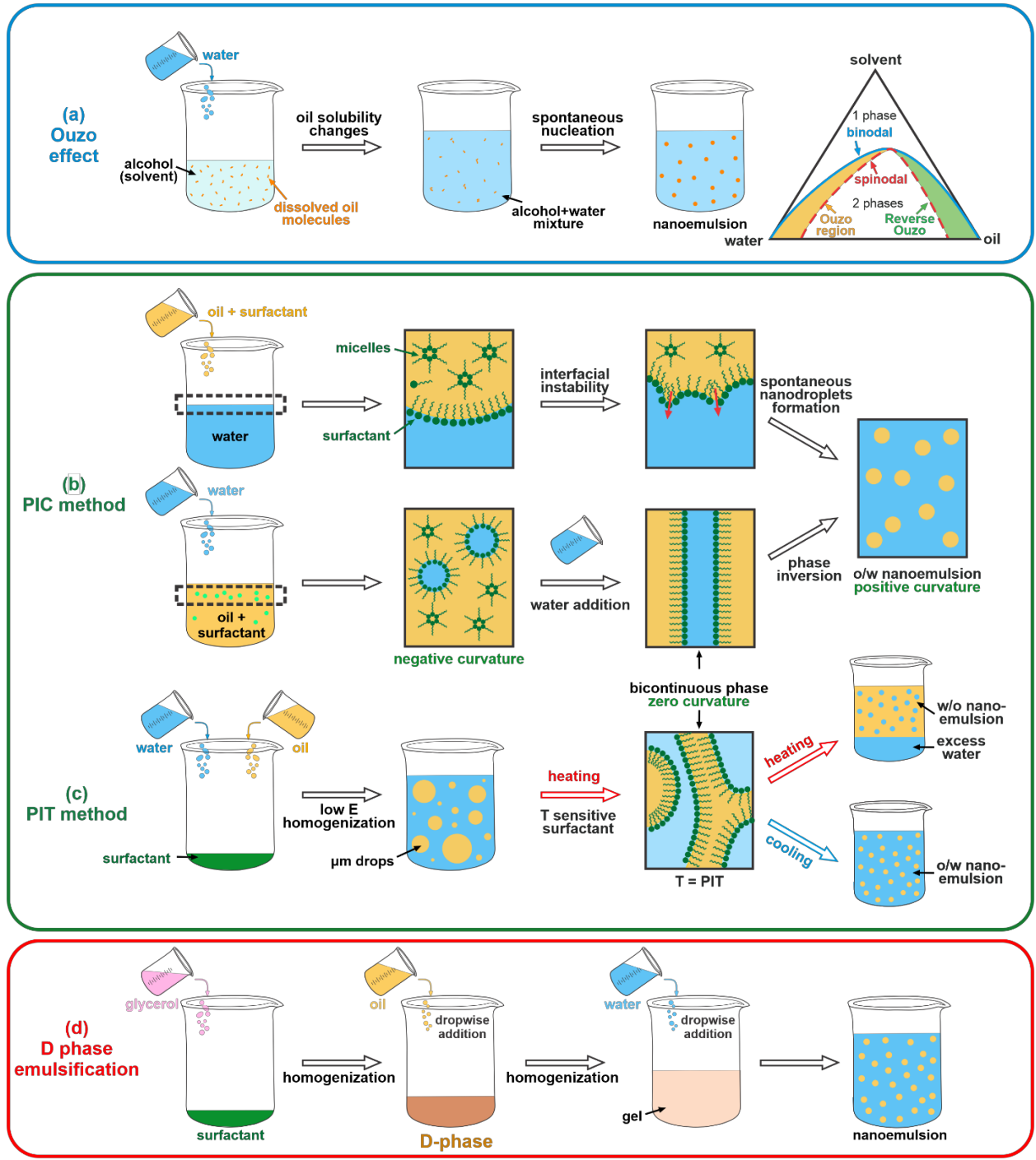

**Figure 2**. Schematic representation of the main low-energy methods for nano-emulsification: (a) via Ouzo effect; (b) PIC method; (c) PIT method and (d) D-phase emulsification.

Besides the water-ethanol-anethole system which has been widely studied, spontaneous emulsification via Ouzo effect has been demonstrated also with several other ternary mixtures, including water-ethanol-eugenol [19], water-ethanol-benzene [20,21], water-ethanol-octane [22,23] and the Italian liquor limoncello [24], thus showing the general applicability of this approach for such mixtures.

The potential applications of the Ouzo effect have been intensively studied in the last several years. The Ouzo effect turns out to be particularly suitable for preparation of polymeric nanoparticles using the nanoprecipitation method [15,25-27**]. In this case, the polymer is usually dissolved in acetone or other appropriate solvent, and large amount of water is added afterwards to the mixture, causing the polymer precipitation under controlled conditions [28,29]. In a recent comprehensive review Yan and co-authors [27**] highlighted the present capabilities of this approach for production of polymeric colloidal particles with different morphologies.

The Ouzo effect was applied successfully also for preparation of other types of nanoparticles: mesoporous silica [30], carbonated strontium hydrohyapatite which is used as biomaterial [31], particles used as drug delivery systems [25], loaded nanoparticles [32] and even liposomes and vesicles [15].

In recent studies, Tan et al. [33*], Raju et al. [34] and Lamb et al. [35] explored the preparation of supraparticles (large clusters of much smaller colloidal particles used in optics, optoelectronics, catalysis, photonics, sensing, drug delivery systems and other applications) by the evaporation of Ouzo droplets. One common problem found in the production of such supraparticles by the drop evaporation technique was the uncontrollable shape of the emerging supraparticles which was associated with the pinning of the contact line of the evaporating droplet, the so-called "coffee-ring effect". Tan et al. [33*] found that this problem can be overcome by using self-lubricating colloidal Ouzo droplets. Along this line, Raju et al. [34] defined some guiding principles to tune the shape and structure of the obtained supraparticles.

*Self-emulsification in the presence of surfactants*

Self-emulsification in the presence of surfactants is often driven by the difference in the chemical potentials of the surfactant molecules in the two immiscible liquids, cf. **Figures 1c** and **1d** [36-46]. In this group of methods the surfactant is initially dissolved in the liquid in which it has lower solubility. After placing this initial surfactant solution (or dispersion if the surfactant is poorly soluble in the respective phase) in contact with the other liquid, a spontaneous process of surfactant transfer occurs which, under appropriate conditions, can induce nanoemulsion formation.

Different triggers have been used to change the surfactant solubility in the contacting liquids – temperature, pH, salt concentration, surfactant structure, surfactant concentration, *etc*. Salager et al. [47,48] defined the concept of "hydrophilic-lypophilic deviation number, HLD," which describes the affinity of the surfactants towards the two immiscible liquids, depending on the experimental conditions (temperature, salt concentration, etc.). This classification has been used to determine what type of emulsion will be formed under given conditions.

Nanoemulsions are often formed in this approach when the boundary between regions with different preferred types of emulsion is passed and the emulsion type is inverted. In other words, initially one of the liquids is dispersed in the other liquid and, after phase inversion, the second liquid forms nanodrops in the first liquid which is transformed into a continuous phase. The recent studies show, however, that the phase inversion is not a strict requirement for nanoemulsification [49**]. Instead, transient bi-continuous phase(s) should be formed at the boundary between the two liquids which is then destabilized to form the nanoemulsion, as explained below.

Two types of phase transitions have been distinguished in the literature. Transitional phase inversion occurs when moving downwards from region of stable W/O emulsions with positive HLD value to a region of stable O/W emulsions with negative HLD value – this can be achieved by changing the formulation parameters (surfactant solubility, pH) or the processing conditions (temperature). In contrast, catastrophic phase inversion occurs by adding an excess of dispersed liquid into an inherently unstable emulsion (initially with very low concentration of the dispersed drops) until the drops in the unstable emulsion coalesce catastrophically and the dispersed phase is transformed into a continuous phase. The latter method relies mostly on the change of the water-to-oil ratio for triggering the inversion [47,48]. These methods have been extensively studied in the last few decades and numerous review papers describing their mechanisms, control parameters and applicability are available [2,5,12,36-48]. Here, we highlight only the main emulsification principles, the role of the intermediate bi-continuous phases as currently understood, and some recent developments in the area.

*Spontaneous emulsification achieved by titration methods*

The spontaneous emulsification achieved by titration, also known as "phase inversion composition, PIC" method was first described by Lin et al. [45,46] nearly 50 years ago. Traditionally, the PIC method involves addition of aqueous phase into the oily phase which contains a pre-dispersed water-soluble surfactant. Various processes and related control parameters have been discussed along the years as important in this method [36,38,43,49**].

Traditionally, the mechanistic explanation of the PIC method is given along the following reasoning. The surfactant molecules, initially dispersed in the oil, are dehydrated. When water is first added to this oil-surfactant mixture, the surfactant head-groups become hydrated which changes the surfactant packing parameter (the area-per-molecule increases) with concomitant change in the preferred (spontaneous) curvature of the surfactant adsorption layers. In this stage, reverse microemulsions are formed preferentially [44]. With the further water addition and enhanced hydration of the surfactant molecules, the spontaneous curvature of the adsorption layers gradually decreases down to zero and lamellar or bicontinuous phase is formed. Eventually, the sign of the curvature of the surfactant adsorption layer is inverted, favoring the formation of oil-in-water droplets which may be of nanometer size, if the conditions are properly tailored. The change in the sign of the preferred curvature and the appearance of ultralow interfacial tension, both occurring at similar compositions of the triple oil-water-surfactant mixture (called "phase-inversion composition" or "PIC"), were discussed in the literature as being crucial for the success of this procedure.

Recent analysis of the available experimental data, however, stresses on the development of a bi-continuous (cubic or sponge) phase as a critical step for successful nanoemulsification in this approach [49**]. Such bi-continuous phases are formed at high surfactant concentration. In the final step of water addition, the oily droplets are assumed to nucleate in the bi-continuous phase through the so-called "templated oil phase separation process" and are quenched in the final oil-in-water nanoemulsion [49**]. In this approach, the size of the oily droplets is explained by the curvature of the surfactant adsorption layer in the bi-continuous phase in the moment of its destruction, which in turn is related to the surfactant-to-oil ratio (SOR) [36,38,49**].

The PIC method is called also a "catastrophic phase inversion" (CPI), "emulsion inversion point" (EIP), and "emulsion phase inversion" (EPI) in the literature. All these names suggest that the actual phase inversion, at which the spontaneous curvature of the surfactant adsorption layer changes its sign, is really important for the nanoemulsification. However, the analysis of Roger [49**] showed that this is not a crucial requirement. He explained that if one starts the experiment with a ternary composition in which the surfactant has already the preferred curvature favoring the oil-in-water emulsions, upon further water addition a nanoemulsion is formed without changing the sign of the spontaneous curvature. This pathway was termed "superior-PIC" (sup-PIC) and it was shown experimentally to lead to most efficient nanoemulsification [49**].

The main mechanism described to cause nanoemulsion formation when the titration is performed in the opposite order, *i.e.* non-polar oily phase is titrated into the aqueous media, is the presence of interfacial turbulence at the interface between the organic and aqueous phases, also called "slightly interfacial disturbance method". This interfacial turbulence leads to oil fragmentation into nano-sized oil droplets by the mechanism of interfacial instability, caused by coupled surfactant transfer across the interface and Marangoni effects, as described by Sternling and Scriven [50]. One should note that similar surfactant transfer and the related nanodroplet formation could occur also in the water titration method during the last stages of water addition – a possibility which is worthy to be investigated more systematically in future studies.

The PIC method is relatively easy to implement from experimental viewpoint, because it does not require any expensive equipment. However it is limited to specific oil-surfactant combinations and all processing conditions affect tremendously the emulsification outcome, including the rate of titration, type of surfactant and oil, surfactant-to-oil ratio (SOR), surfactant-to-water ratio (SWR), processing temperature, salinity, pH and stirring speed [37,40]. Although hundreds of papers explore the applicability of this method to various systems, the requirements about the oil and surfactant type which allow the formation of nanosized droplets are not entirely understood [37,40] and (almost) no general design rules are currently available.

The surfactants which are used most frequently in the PIC method are the water-soluble polyoxyethylene sorbitan monoalkylates (known as polysorbates or Tweens) and their oil-soluble analogues - sorbitan monoalkylates (Spans) [51-58]. As an example, Barzegar et al. [59] produced 20 wt. % of peppermint emulsions with drops between 50 and 190 nm using Tween surfactants. In this case, the organic phase was titrated into the aqueous media for a period of 10 min and the whole mixture was stirred at 700 rpm for a period of three hours. Best results were achieved with Tween 80 surfactant, but the required surfactant amount was between 0.5 and 2 times the amount of the oil, i.e. 10 to 40 wt. % surfactant.

In another study, Ostertag et al. [51*] compared the emulsification outcome for medium-chain triglyceride oil (MCT), orange and limonene flavor oils, and different long-chain triglyceride oils extracted from olives, grape seeds, sesame, peanuts and canola plants by using Tween 80 as surfactant. The smallest droplets were obtained with MCT oil ($d_{32} \approx 140$ nm), followed by flavor oils with $d_{32} \approx 300$ nm, whereas the average drop size for the other oils exceeded 500 nm or even 1000 nm. The authors could not find a clear explanation about this trend when analyzing the oil properties (mass density, viscosity, interfacial tension of buffer/oil interface). On the other hand, they did not report the interfacial tensions of these oils in the presence of surfactant which may be the governing factor in the emulsification process.

In the last years, the drive to exchange the synthetic surfactants by surfactants of natural origin ("green" surfactants) resulted in several studies exploring the spontaneous emulsification with phospholipids, sugar esters and other "greener" surfactants.

Promising results were demonstrated by Luo and co-authors [60] who prepared 160 to 200 nm clove bud oil emulsions via self-emulsification using whey protein concentrate for stabilization. However, alkaline solution was used for initial dissolution of the oil which could lead to some chemical transformations in the oil and in the protein stabilizer.

Ariyaprakai and co-authors [61] studied the spontaneous formation of flavor oil emulsions, stabilized by sucrose esters. The smallest mean diameter in the system was around 200 nm, obtained with 10 wt. % orange oil, 7.5 wt. % sucrose esters (S1670 or P1670) and small amount of ethanol, whereas the droplets were much bigger at lower surfactant concentrations. The obtained drops of peppermint oil in the absence of alcohol were significantly larger (1.5 to 4 μm) for all investigated compositions.

Sugar ester (glucose monooleate) and cremophor EL (polyoxyethyleneglycerol triricinoleate 35) were studied in a combination by Ishak et al. [62*] as stabilizers for vegetable oils (coconut oil, sunflower oil, palm oil, etc.). Drops with size as small as 40 nm were produced with 10 wt. % oil and 15 wt. % surfactant mixture for coconut oil and sunflower oil, whereas larger drops were obtained with the other studied oils. The experiments with FTIR, SAXS and observations in polarized light showed that coconut oil-surfactant-water and sunflower oil-surfactant-water system formed lamellar/bicontinuous phases at optimal composition, whereas the other oils (olive, palm, etc.) did not form such phases.

Concluding, the spontaneous emulsification method at constant temperature has been mainly explored in the recent years for preparation of nanoemulsions with essential oils which are known to have antimicrobial and antioxidant activity [54,58-60,63-65]. Potential applications related to enhanced oil recovery have been also studied [66,67].

*Phase inversion temperature method (PIT)*

The PIT method for nanoemulsification, demonstrated for the first time by Shinoda and Saito [68], is based on changes of the surfactant properties with temperature. The surfactants containing polyoxyethylene units are most widely studied, because their hydration degree and spontaneous curvature change significantly with temperature. At low temperature, the water-soluble ethoxylated surfactants have large spontaneous curvature and form direct micelles, whereas at temperatures higher than the phase inversion temperature (PIT, called also "HLB temperature"), the surfactant molecules adopt negative spontaneous curvature and form reversed micelles. Around PIT, the average spontaneous curvature is close to zero and, therefore, lamellar or bicontinuous structures are formed. The interfacial tension is extremely low at this temperature, favoring both the emulsification and drop coalescence [69*]. Many factors have been shown to affect the phase inversion temperature, including SOR, WOR, oil type, presence of electrolytes and other additives [70].

The understanding of PIT mechanism has evolved along the years and it is currently accepted that the main structures responsible for the nanodroplet formation are those with zero spontaneous curvature [49**,69*,71,72]. Morales and co-authors showed that it is of major importance that the oil and surfactant coexist in the same phase at the PIT, and there could be some excess of water in the system [73,74]. It was argued that the PIT method uses the surfactant more efficiently than the PIC method, because the preferred curvature of the produced droplets matches the drop curvature and no additional surfactant remains unused in the aqueous phase [49**,69*,71]. However, various definitions for the surfactant efficiency can be used – in the practical applications, the weight efficiency is preferred, viz. obtaining drops with similar size at lower SOR is more efficient.

The procedure for nanoemulsion formation via the PIT method includes preparation of a coarse oil-in-water emulsion at low temperature, for which purpose gentle stirring is usually applied. Afterwards, the temperature is increased close to the PIT or above it, in order to facilitate the phase inversion process. The sample is kept at the PIT for a certain period to allow the formation of bicontinuous or lamellar liquid crystalline phases. Then the system is rapidly cooled upon continuous stirring. Thus, kinetically stable oil nanodroplets are formed via thermal disruption of the zero curvature phase, due to changes in the surfactant hydration properties and in the related spontaneous curvature of the surfactant adsorption layers [69*]. The three-phase diagrams water-surfactant-oil (fish diagrams), as a function of temperature, were particularly useful to determine the optimal condition for PIT emulsification.

Roger and co-authors demonstrated that, similarly to the PIC method, the PIT method may be realized at sub-PIT levels, viz. at lower temperatures than the exact PIT [49**,71]. The process is best realized at a few degrees below the phase inversion temperature when bicontinuous microemulsion coexists with an excess of oil phase. Upon gentle stirring, oil-in-water microemulsion appeared as a metastable structure, which transformed into kinetically stable (quenched) nanoemulsion after cooling. The drop sizes were similar to or smaller than those obtained after the conventional PIT method in which the temperature is raised above the actual phase-inversion temperature [71].

In recent years, PIT method has been successfully used to produce essential oil emulsions [75-77]. Also, Lemahieu et al. [78*] used PIT-related technique for determination of the equivalent alkane carbon number (EACN) of crude oil, needed to optimize the surfactant formulation for enhanced oil recovery. This approach was applied by Dario et al. [79] and McClements at al. [80] to produce nanoemulsions in the presence of cationic surfactant.

The PIT emulsification technique has been used also for preparation of nanoemulsions containing monomer droplets, as precursors for polymeric latex particles after additional polymerization step [81-83*]. In these systems, the PIT temperature must be higher than the polymerization temperature to ensure that the prepared nanoemulsions remain stable prior to polymerization. Sasaki and co-authors showed that monomodal latexes with tiny particulates are produced when the polymerization temperature is relatively close to the PIT, whereas at lower polymerization temperature the latexes had bimodal size distribution with relatively large particles being also present [83*]. Boscan and co-authors prepared by this method ca. 100 nm polylauryl methacrylate particles with 33 % weight fraction starting from emulsion stabilized by 4.5 wt. % surfactant [81].

The production of inverse, water-in-oil emulsions by the PIT method has been also achieved [84]. In this case, the oil-surfactant-water mixture is maintained in the phase inversion zone, while being diluted with additional oily phase, thus combining the PIT and PIC approaches.

Concluding, droplets as small as *ca.* 20 nm have been produced by the PIT method. As any other method of emulsification, this method also has its limitations. High concentrations of surfactants are usually needed (SOR = 1 to 3). The requirement for phase inversion limits the method to surfactants which are able to change their curvature upon temperature variations. Very importantly, the method is not suitable for temperature-sensitive compounds (e.g. some drugs).

*Phase inversion pH method (PIpH)*

Wang et al. [85*] proposed an alternative method for low-energy nanoemulsification which uses phase inversion, induced by pH changes in the aqueous phase (PIpH). For that purpose, pH responsive surfactants based on amine-carboxylic acid complexes were synthesized. Starting from pH of 8.6 at which O/W was present, the authors induced phase inversion to W/O emulsion by decreasing pH down to 8.3 and then a second phase inversion back to O/W nanoemulsion was induced by subsequent pH increase up to 9. In this way, 80 nm liquid paraffin droplets were produced from initial coarse emulsions at WOR = 0.5 and surfactant concentration of *ca.* 5-8 wt. %. This study is intriguing and seems important, although it remains unclear how the reported interfacial tensions < 0.01 mN/m (down to 0.0018 mN/m) were measured [85*] by the pendant drop method which is known to be applicable to interfacial tensions higher than *ca.* 0.5 mN/m.

*D-phase emulsification*

The so-called "D-phase" emulsification technique is a version of the phase-inversion methods (PIC and PIT) which allows nanoemulsion formation at lower surfactant concentrations (e.g. 2-3 wt. % *vs.* 10 wt. % in the PI methods) at the expense of using an additional component in the system. The D-phase method has been introduced first by Sagitani et al. in 1983 [86] and studied briefly by his scientific group in the following decade [87-91]. The method became popular again in the last 5 years with several research groups exploring its applicability to various systems [92-97].

The method uses surfactant (mainly polyoxyethylene alkyl ethers or polyoxyethylene sorbitan esters), oil, water and a fourth component (usually alkyl polyol) to produce kinetically stable nanoemulsions. Different sugars can play the role of polyols as well [98]. The role of the additional component is to alter the hydration properties of the ethylene oxide groups in the surfactant heads and, therefore, to change the surfactant HLB value and the related cloud point. Two different mechanisms of this method have been proposed – a fraction of the water molecules around the EO groups may be replaced by the polyol molecules or the properties of the aqueous phase could be modified by the polyol, thus decreasing the water content around the surfactant headgroups [90,99].

In practice, the method is realized in several steps. First, the surfactant is mixed with the polyol, possibly in the presence of small amount of water (e.g. 2-3 wt. % water in the mixture). The homogeneous mixture prepared in this way is called "D-phase". Afterwards, the oil is added in a dropwise manner upon continuous stirring, usually on a magnetic stirrer at 300-500 rpm. The result is an oil-in-surfactant dispersion, called "O/D phase", which often has relatively high viscosity. After subsequent stirring for *ca.* 20-30 min, the obtained dispersion is diluted with slowly added water. The process may be conducted at elevated temperature to optimize the surfactant hydrophobicity.

Following this procedure, Endo and Sagitani produced 500-850 nm olive oil-in-water nanoemulsions stabilized by 2 wt. % polyoxyethylene (20) oleyl ether in the presence of *ca.* 1 wt. % polyol (1,3 – butanediol, propylene glycol, glycerol or PEG 400). Smallest drops were obtained in the presence of glycerol ($d \approx 500$ nm) [88].

Zhang and co-authors studied the effect of the oil type for the D-phase emulsification technique, exploring different linear and branched alkanes and several triglyceride oils [93]. These authors concluded that the alkanes with lower molecular mass are more likely to produce smaller nanoemulsion droplets. However, these emulsions were less stable upon storage when compared to the emulsions prepared with alkanes of higher molecular mass.

In one series of studies, this method was used for preparation of a series of nanoemulsions with saturated linear alkanes [92,95,96]. Even-numbered alkanes with chain lengths between 14 and 18 C-atoms were studied, because they melt in the range between 5 and 25°C. These alkanes have relatively high latent heat and, therefore, their emulsions are explored for energy storage and energy transport applications [100-102]. Using this low-energy emulsification method, the researchers prepared emulsions with drop size of ca. 200 to 600 nm, at relatively high oil fraction of 10 to 40 wt. % using relatively low concentrations of surfactant and 1,3-butandiol 992,95,96].

Recently, Yukuyama and co-authors [97] used D-phase emulsification to produce MCT nanodroplets, loaded with flubendazol drug. In this study, emulsions with drop sizes down to 35 nm were prepared at 60 wt. % oil content using 3 wt. % surfactant and 3 wt. % glycerin only. These emulsions were reported to remain stable for a period of 3 months.

As seen from this short summary, the D-phase emulsification method has high potential for production of concentrated emulsions at lower surfactant levels. However, there are many open questions to be answered. For example, what is the exact structure of the phases obtained in the different emulsification stages and how they change along the process; what is the mechanism of nanodroplet formation; what are the limitations of the process with respect to the ingredients and can we observe the same process with natural surfactants? As with the other low-energy methods, such a deeper understanding would allow the researchers to define general design rules for application and optimization of this intriguing and potentially very useful method.

## 3. Recent advances in the self-emulsification techniques.

Several novel self-emulsification techniques were proposed recently. We discuss three of them which seem to us most promising for further development and applications: the MAGIQ method (monodisperse nanodroplet generation in quenched hydrothermal solutions), spontaneous emulsification driven by polymorphic phase transitions in lipids, and nanoemulsification when multiconnected micellar phase is placed in contact with hydrophobic fragrance phase.

### *MAGIQ method*

Deguchi and co-authors discovered a new method for nanoemulsion formation [103,104**] which is based on the altered physicochemical properties of water upon increase of pressure and temperature close to its gas-liquid critical point, $T_c$ = 374°C and $p$ = 22.1 MPa. The supercritical water (SCW) has a very low dielectric constant of $\varepsilon \approx 2$ which suggests that it may be able to mix freely with many hydrophobic oils [105-107]. The MAGIQ process is based on this peculiar property of the SCW and it is realized in two steps: first, the SCW and hydrocarbon are mixed at elevated temperature and pressure into hydrothermal homogeneous solution which is afterwards quenched by a rapid and deep temperature drop (≈ 200°C/s cooling rate) via injection into aqueous surfactant solution, see **Figure 3a**. Emulsion yield of 10 ml/min at oil volume fractions of up to 10 vol. % was demonstrated in a lab-based equipment [103,104**].

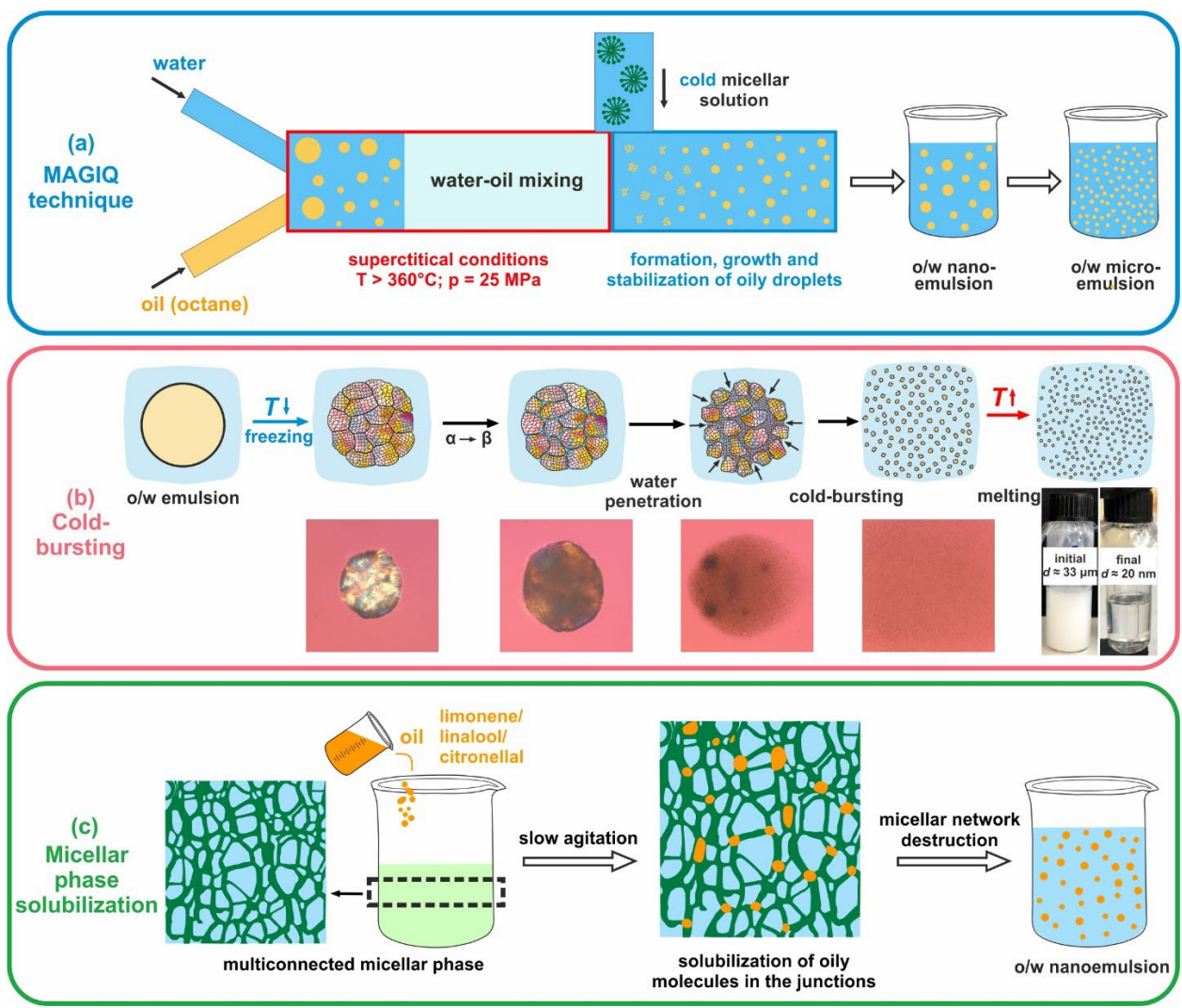

**Figure 3.** Schematic representation of three novel self-emulsification methods discovered in the last years. (a) MAGIQ technique [103,104**]; (b) Cold-bursting method [113**,114] and (c) Nanoemulsion formation by micellar phase solubilization and fragmentation [119**].

It was found [104**] that the mechanism of drop generation involves initial oil phase separation from the binary oil-SCW solution upon cooling and quenching, followed by surfactant adsorption on the newly formed oily drops. In contrast to the other low-energy methods, utilizing the changes in the surfactant properties upon change of the temperature and/or composition of the emulsion, the surfactant used in MAGIQ method is needed only for stabilization of the oily drops against coalescence. It was proposed that many different surfactants can be applied in this process, although polyoxyethylene-10 oleyl ether (Brij 97) was only used in the reported studies.

Beside the potential technological applications, this method is interesting also in the context of the studies about the origin of life. An increased scientific interest arose in the last years about the possibility that the life may have started at the deep-sea hydrothermal vents [108-112]. The conditions in these vents (temperature, pressure) resemble those used in the MAGIQ method and could drastically change the behavior of the known hydrophobic and amphiphilic substances [103,104**].

### *Spontaneous cold-bursting*

The polymorphic phase transitions in lipid phases is used in another novel approach for spontaneous fragmentation, called "cold-bursting method" [113**,114]. This method produces submicrometer particles from large lipid drops (initial size of up to 100 μm) by simply cooling and heating of the lipid dispersion around the melting temperature of the dispersed oil, $T_m$, or upon prolonged storage of the dispersion at temperature close to, but lower than $T_m$. Nanodroplets and nanoparticles with diameter down to 20 nm were formed in such emulsion systems after few consecutive cooling-heating cycles. Below, we review briefly the main features and the underlying mechanisms of the cold-bursting process.

Crystalline triglycerides are known to arrange in three polymorphic modifications: α, β' and β, ordered here from α-phase with lowest melting temperature to β-phase with the highest thermodynamic stability and the highest melting temperature [115]. This polymorphism is of monotropic type, *i.e.* the β polymorph has the lowest Gibbs energy in the whole temperature range of solid existence. Hence, once the β phase is formed, the other two phases (α and β') do not form, unless the system is melted and recrystallized again. On the other hand, upon cooling, the triglyceride molecules usually arrange first in the least-stable α phase as the nucleation barrier for this transition is the lowest. Afterwards, this α phase transforms into one of the more stable polymorphs. This transition can occur in two different modes: either it can be melt-mediated – if the temperature of the sample is increased above the melting temperature of the α phase but below the melting temperatures of the β' or β phase, or it can be a solid-state transition which is observed upon prolonged storage of the sample at temperature below the melting temperature of the α phase.

On particular feature of these lipid systems is that the α-phase has significantly lower mass density than the β-phase – therefore, the crystalline domains shrink in the transition α→β. The local contraction of these crystalline domains causes formation of nanovoids at the grain boundaries in the frozen triglyceride particles and it is associated with the so-called "negative pressure effect" [116-118].

Nanoparticles are obtained in the cold-burst method, starting from emulsions of much bigger lipid drops in appropriate surfactant solutions, see **Figure 3b**. Upon cooling, the lipid drops freeze in a polycrystalline structure, containing domains of α-phase. Upon prolonged subsequent storage at a temperature below the melting temperature of the α-phase or upon heating above this temperature, α→β transition occurs, causing the formation of a three-dimensional structure of β-domains separated by nanopores in the big lipid particles. Due to the negative pressure effect, the outer continuous phase (aqueous surfactant solution) is sucked into the particle interior. A related increase of the particle volume is observed, followed by fragmentation of the original large lipid particles into thousands or millions of β-crystalline individual nanoparticles.

Systematic study [113**] showed that the cold-bursting is observed in the systems for which the three phase contact angle formed between the solid triglyceride, the aqueous surfactant solution and air is ≤ 50°, *i.e.* when the triglyceride is wetted well by the aqueous surfactant solution. Thus, appropriate oil-surfactant pairs were identified which lead to efficient nanoparticle formation by this method.

The most efficient temperature protocol included rapid cooling followed by slow heating, because the higher cooling rate ensures higher in number and smaller in size crystalline domains, whereas the slower heating is needed to provide sufficiently long time for penetration of the aqueous phase into the lipid particles and subsequent particle disintegration. Nanoemulsions containing up to 20 vol. % oil and drops of ca. 100 nm or smaller were produced by this method, without any mechanical input to the system [113**].

In a following study [114] exploring the applicability of the cold-bursting method to complex in composition lipid particles, several additional factors were found to affect in a non-trivial way the method efficiency. Understanding the importance of these additional factors and the related mechanisms allowed the authors to optimize in a rational way the conditions for efficient cold-bursting in various lipid systems, some of them used widely in real applications.

So far, the cold-bursting process has been successfully applied to more than a hundred oil-surfactant combinations. The oils include monoacid triglycerides with various chain-lengths (between $C_{12}$ and $C_{18}$); natural triglyceride mixtures, such as coconut oil, palm kernel oil, cocoa butter; diglyceride oils – dilaurin and Precirol ATO 5; different long-chain alkanes and also phospholipids (DPPC). Wide range of nonionic surfactants has been found to be efficient, as well as various cationic and anionic surfactants, and their mixtures. The process is efficient also for triglyceride particles loaded with active substances of relatively high concentrations, e.g. with fenofibrate drug loaded up to 30 wt. % in the lipid particles [114].

*Nanoemulsions formation by multiconnected micellar phase*

In one of his last and very interesting publications, P. Kralchevsky and co-authors described another method for low energy nanoemulsification [119**]. They obtained 100-200 nm emulsion droplets by placing different fragrance oils (limonene, citronellal and linalool) in contact with saturated micellar network of interconnected branched micelles.

The phase of saturated micellar network, observed and theoretically predicted also by other authors [120,121], was formed when appropriate amount of divalent cations was added into the mixed surfactant solution of SLES+CAPB (sodium lauryl ether sulfate + cocamidopropyl betaine). Upon storage, these surfactant solutions separated into two phases – lower phase containing the main fraction of the surfactant in the form of connected branched micelles and an upper phase containing mainly water, **Figure 3c**. The total surfactant concentration in these experiments was varied between 0.2 and 5 wt. % and the salt concentration – up to 100 mM. The authors found that most efficient phase formation of saturated micellar network was observed with $Mg^{2+}$ cations as compared to $Zn^{2+}$ and $Ca^{2+}$.

Nanodroplet formation was observed with several oils when 1 wt. % surfactant solution + 30 mM $MgCl_2$ was used. When the oils were added to the micellar phase and stirred for an hour, nanodroplets as small as 100 nm were observed. The method was realized at surfactant-to-oil ratios comparable to those in the other spontaneous emulsification methods, SOR $\approx$ 1.36, 1.74 and 3.44 for limonene, linalool and citronellal, respectively [119**]. These results were obtained at relatively low oil weight fractions, between 0.74 and 0.29 wt. %, and further studies are needed to check the method limits with respect to the oil concentration.

The authors suggested that the nanoemulsion formation is related to the solubilization of the oil molecules within the junctions of the micellar network. These junctions swell as a result of the oil solubilization and separate as nanodroplets while the micellar network is destroyed after some critical junction size is reached [119**]. Further investigations would help to verify this proposed mechanism and to check whether bi-continuous structures, similar to those reported in the phase-inversion methods, are involved in the intermediate stages of this process.

## 4. Self-emulsifying drug-delivery systems

Lipid-based formulations (LBF) have been studied for many years in the pharmaceutical context and the subset of self-emulsifying drug-delivery systems (SEDDS) was defined in the 1980s, more than 40 years ago [122]. This last class of systems was reinvented in the last decade, mostly due to the flow of encouraging data from animal studies and clinical trials, demonstrating the significant benefits in terms of increased oral bioavailability and reduced interindividual variability in exposure, offered by this approach. Detailed description of the materials used to prepare LBF, including self-dispersing formulations, are available in the book by D. Hauss [123]. Below we address the SEDDS approach in particular, while the reader is referred to a recent general review [124] for the other LBF technologies.

The use of self-emulsifying formulations for the delivery of active pharmaceutical ingredients to human patients bears several specific aspects. As most SEDDS are intended for oral intake, they are prepared as lipid solutions that self-emulsify upon contact with the fluids in the human gastrointestinal tract (GIT). Therein, the SEDDS encounters a complex medium, which contains endogenous surfactants (bile salts, phospholipids) and digestive enzymes (lipases, proteases, carboxylesterases etc.) which can affect significantly the self-emulsification process and the stability of the emulsions formed. In addition, one must take into account the formulation constraints imposed by the safety requirements – only biocompatible, low-toxicity compounds can be used for SEDDS preparation.

The development of SEDDS is further complicated by the fact that it requires knowledge in several disciplines: physical chemistry of emulsions (to facilitate formulation), pharmacokinetics (to understand the clinical implications) and human physiology (to bridge formulation chemistry with the clinical impact). Various biopharmaceutical aspects of the SEDDS performance have been addressed in a recent special issue [122]. Therefore, we focus our discussion below on the physicochemical behaviour of SEDDS at the conditions of the GIT.

The counterintuitive terminology used in the chemical sciences (*e.g.* nanoemulsions vs. microemulsions, see Section 1) has left its mark in pharmaceutics where numerous formulations are labelled as "self-microemulsifying", "self-nanoemulsifying" or just "self-emulsifying" without proper justification. Usually, there is no data to support the claim that the system is truly thermodynamically stable after dispersion (a microemulsion), or that it is kinetically stabilized (a nanoemulsion). Instead, the various "self-emulsifying" terms are used to denote all lipid solutions which produce homogenous, slightly opaque or transparent emulsions after addition into aqueous media via mild agitation. This issue was addressed first by Anton & Vandamme [125] and discussed further by Niederquell & Kuentz [126], who performed dedicated stability experiments and suggested specific size boundaries for the different classes. Based on experimental data and theoretical considerations, the authors suggested that true microemulsions (or swollen micelles) are systems that yield an aggregate diameter < 15 nm upon dispersion, whereas kinetically stabilized emulsions are obtained when the diameter is > 100 nm [126]. For the systems with intermediate size, between 15 and 100 nm, the authors concluded that thermodynamic stability can only be proven using appropriate stability experiments. Unfortunately, the drop size for most developed SEDDS falls exactly into this "intermediate" category [127-129] where the thermodynamic stability cannot be assumed from first principles. In the current review, the term "SEDDS" (self-emulsifying drug delivery systems) is used as a surrogate for both "self-microemulsifying" and "self-nanoemulsifying" formulations, in order to illustrate their propensity to form uniform dispersions upon dilution and mild agitation, without bearing any information regarding their thermodynamic stability.

Two general questions always appear in the SEDDS studies: (1) what are the mechanisms of spontaneous emulsification and (2) how can we control the self-emulsifying properties of a lipid formulation? The answers are related to the chemical composition of the formulations and their interactions with the GIT fluids (especially with the enzymes in these fluids).

To structure the discussion of the SEDDS formulations, we use the LBF classification system suggested by Pouton [130] where they appear as type II or type III. According to this classification, LBF of type II contain 20-60 % surfactants which provide their self-emulsifying functionality, which is further enhanced by the introduction of co-solvents (up to 50 %) in type III

LBF. Therefore, SEDDS of type II and III LBF, which are of interest in the context of the current review, are composed by oils (triglycerides or their mixtures with mono- and diglycerides), surfactants (mainly ethoxylated species, such as polysorbates, ethoxylated castor oil etc.) and cosolvents (ethanol, PEG-400, polyols etc.).

Type IV LBF are also defined in this classification and they consist of a mixture of solvents and surfactants only (no triglycerides or other oils present). Obviously, these latter systems exhibit the best self-emulsifying properties. However, as noted by Niederquell & Kuentz [126], they should not be discussed in the context of emulsions and SEDDS at all, due to the absence of water-immiscible phase.

The composition of SEDDS can be used to draw parallels with the general self-emulsification approaches and mechanisms discussed in the previous sections. The high surfactant concentration in SEDDS suggests a compositional similarity with the phase-inversion composition (PIC) method discussed above. On the other hand, the cosolvents used in some SEDDS resemble the formulations described in the D-phase method. Hence, the composition of SEDDS is expected to activate the same self-emulsification mechanisms as discussed in Section 2 above – an assumption which has been confirmed by recent structural studies, as explained below.

When a SEDDS comes in contact with an aqueous phase, the isotropic, surfactant-rich oil phase is impregnated by water, resulting in the transient formation of lyotropic liquid crystalline phases, as evidenced my small-angle X-ray scattering (SAXS) measurements [131]. Further dilution can induce the formation of a hexagonal phase (depending on the surfactant used) and finally – microemulsion. Coarse-grain molecular dynamics simulations confirm the phase transitions expected upon dilution of lipid formulations with water and hint to the formation of inverse micellar phase prior to the lamellar liquid-crystalline phase [132], **Figure 4a**.

In another recent study, space- and time-resolved SAXS experiments were used to explore the phases obtained upon dilution of a lipid formulation with a composition similar to SEDDS in a range of biorelevant fluids [133*]. By reconstruction of the 2D spatial maps of the oil-water mixture, the authors showed the formation of liquid crystalline-rich regions containing inverse bicontinuous Im3m and Pn3m cubic phases, demonstrating that this experimental approach can be used to gain more insight on the complex phase behaviour of SEDDS formulations.

The impact of the aqueous media composition on the self-emulsification has not been studied systematically. However, results from drop size measurements showed that the biorelevant surfactants, such as bile salts and phospholipids, led to smaller emulsion drops, when compared to the plain buffer [134]. The loading of drug in the SEDDS had limited effect on the drop size after dispersion of formulations containing low fraction of lipids (< 10 %), whereas it increased slightly the drop size at higher lipid concentrations (> 20 %) [134].

Self-emulsification creates an emulsion with large surface area, which is prone to digestion by the enzymes in the GIT. The chemical transformation of the triglycerides to their more polar derivatives (monoglycerides, fatty acids) may result in a digestion-driven phase transition of the lipid droplets – a specific process for this type of systems which serves as both significant complication and an opportunity for system control, thus creating a novel avenue for valuable research in this research field.

Results in this area were obtained either by scoping equilibrium properties of the system via phase diagrams, dynamic light scattering and polarized light microscopy [135], or by time-resolved SAXS coupled with *in vitro* digestion setup which allows one to follow the process dynamics [136-139]. For example, a phase diagram of oleic acid-based systems (mimicking digested LBF) indicates that an inverse micellar system is formed at high lipid concentrations (corresponding to the regions next to the oil drop surface during enzyme lipolysis), which turns into lamellar liquid crystalline and cubic phases upon dilution with water, **Figure 4b** [135]. Further dilution leads to the formation of vesicles and/or micelles, depending on the bile salt concentration. This general trend was confirmed by time-resolved SAXS for the digestion of triolein, which showed the formation of micellar cubic, inverse hexagonal, and bicontinuous cubic liquid-crystalline droplets, finally transfoermed into vesicles [136]. Interestingly, liquid-crystalline structures were not produced during the digestion of medium-chain triglycerides (C8-C10), where vesicles were observed as quickly as 2 min after the start of digestion [136,139].

As another example, the digestion of a SEDDS formulation containing surfactant of high concentration (30 % Cremophor RH40) and solvent (10 % ethanol), caused rapid formation of a hexagonal phase, which coexisted with a lamellar phase until the end of the *in vitro* digestion experiment [136]. Although lipid digestion reached a plateau at the end of the experiment, it was not clear whether a complete lipolysis was obtained (as would be the case *in vivo*), causing some uncertainty about the implications of the obtained results. Further experiments by the group of Müllertz with the same SEDDS showed that the lamellar phase was formed during the dispersion of the formulation in the bile salt and phospholipid-containing aqueous phase, whereas the transition to a hexagonal phase was induced by the digestion [138]. In this case, the authors reported a degree of digestion between 60 and 100 % and suggested that the hexagonal phase dominated over the lamellar liquid crystalline phase when the SEDDS was completely digested.

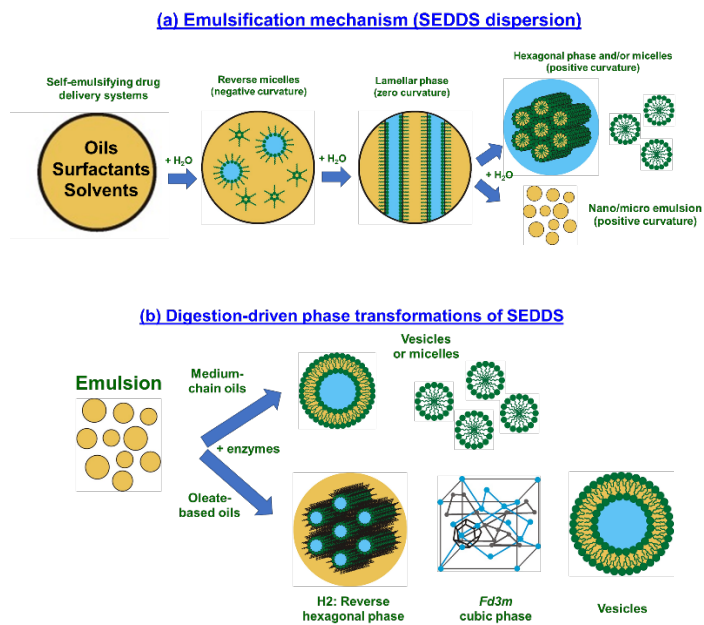

**Figure 4**. Emulsification mechanisms of typical SEDDS formulations and digestion-driven phase transformations in presence of lipolytic enzymes.

Therefore, although more complex in composition, the behaviour of the digested SEDDS is similar to that of the key compounds that are produced during lipolysis of triglycerides. For example, one of the main lipolysis products of oleic acid-based lipids (such as soybean oil, olive oil etc.) is glycerol monooleate (GMO) – a substance with capabilities to form a variety of lyotropic liquid-crystalline phases which are studied extensively in literature [140,141].

A new trend in the development of SEDDS is the drive to prepare solid formulations using appropriate solid carriers with large surface area which can adsorb the oily solution of the drug, thus forming a powder that can be used as a dosage form directly (in sachets) or after tableting [142-144]. A recent study by the group of Boyd showed that a lamellar liquid crystalline phase is obtained during dispersion and digestion of a solid SEDDS formulation based on Gelucire 44/14 (pegylated lauryl-glycerides) [145]. For the monoolein-based SEDDS, the authors observed a quick formation of a micellar/vesicular system, which was attributed to very fast solubilization of the cubic structures (typical for monoolein) by the high concentration of bile salts in the dispersing medium. Hence, the results indicate that the solid carrier does not change dramatically the behaviour of SEDDS, compared to the standard liquid formulations. Note that most studies on solid SEDDS are focused on the biopharmaceutical aspects of the formulations, thus leaving largely unexplored the intricate mechanisms of self-emulsification and digestion-driven phase transitions of these complex systems.

Another promising new area in the application of self-emulsifying formulations with non-trivial properties in bio-relevant media are the charge-reversal systems developed by the group of Bernkop-Schnürch [146*]. In this method, the surface charge of the droplets changes from negative to positive as a result of an enzymatic reaction which releases phosphate anions from the head-groups of the surfactant stabilizing the lipid particles. The major aim of this charge reversal is to bypass the anionic obstacles on the way of the lipid particles to their target cells and to achieve higher cellular uptake at the target site.

One can conclude that the SEDDS systems are typically much more complex in composition, due to the specific regulatory requirements and the enzyme activity in the GIT. Nevertheless, the main physicochemical processes and control factors for self-emulsification have been confirmed so far to be similar to those described in Section 2 with the simpler chemical systems, thus giving a promise for cross-fertilization of ideas between these two very different fields of research.

**5. Outlook**.

The interest in the low energy self-emulsification techniques will further increase in the coming years driven by the transition to "greener" chemical technologies and by the quest for efficient and robust self-emulsifying formulations for oral drug delivery.

The overview of the literature presented above reveals that the "classical" self-emulsification techniques hold a great promise which is, however, still underexplored. The main problem is that the detailed structural studies which reveal the specific mechanisms in the various classes of systems are rather scarce which precludes the possibility (so far) to define clear general design rules, allowing to optimize rapidly the processing conditions for each new chemical system of practical interest. Illustrative examples of this last statement are:

(1) The lack of reliable information about the mechanisms of the D-emulsification method which appears to be more versatile and potentially very useful. Without having a clear mechanistic picture on how the various polyol molecules affect the self-emulsification process it remains elusive to define appropriate control parameters and to use them for rational process optimization;

(2) The lack of guiding rules on how to decrease the surfactant-to-oil ratio (SOR) when producing nanoemulsions by the classical self-emulsification methods. The experimental studies have shown that the choice of surfactant and the processing conditions depend strongly on the type of oil (hydrocarbon, triglyceride or essential oil) and many other factors, some of which being difficult to control. The same studies show that very often the method of trial-and-error is still of major importance when optimizing these systems;

(3) The scarcity of systematic studies to reveal the detailed mechanisms of nanoemulsion formation in self-emulsifying drug delivery systems and to relate these mechanisms to the systems encountered in the chemical technologies which are better understood (due to their chemical simplicity). Such a comparison is expected to boost the cross-fertilization of ideas and specific technical solutions between these two application areas.

In parallel, the methods reviewed in Section 3 prove that conceptually new approaches can still be discovered. All these new approaches rely on the use of phase transitions and/or the contact of non-equilibrated phases, however, combining them in different mechanistic modes and at different chemical compositions as compared to the "classical" methods known for decades. These new approaches have their specific advantages and limitations. For example, the cold-bursting method is particularly appropriate for long-chain triglycerides for which the other known methods usually encounter difficulties. This method is also of particular interest in systems containing temperature-sensitive compounds, because the processing temperature is always kept around and/or below the melting temperature of the lipid phase. On the other hand this method is not applicable to essential oils (for which some of the other methods are very appropriate), because their melting temperature is very low.

With the advance of our understanding on both the classical and the new approaches, one could expect a clear "map of the methods" to emerge which relates the requirements of the specific chemical or pharmaceutical application with the most promising low-energy methods for self-emulsification. We do hope that the current review is a right (though relatively small) step into this promising direction of research.

**Acknowledgements:**

The study received funding from the Bulgarian Ministry of Education and Science, under the National Research Program "VIHREN", project ROTA-Active (no. KP-06-DV-4/16.12.2019).

**Contributions:**

D.C. – conceptualization; investigation; visualization; writing – original draft

Z.V. – investigation; visualization; writing – original draft (pharmaceutical drug-delivery systems)

S.Tc. – conceptualization; writing – review and editing

N.D. – conceptualization; methodology; supervision; writing – review and editing; funding acquisition